\documentclass[reprint,amssymb, amsmath, aip,cha,twocolumn]{revtex4-1}
\usepackage{graphicx}
\usepackage[caption = false]{subfig}
\usepackage{color}
\usepackage{epsfig}
\usepackage{ifpdf}
\usepackage{url}%
\usepackage{amsmath}
\usepackage[colorlinks=true,linkcolor=blue]{hyperref}%
\usepackage{cleveref}
\usepackage{bm}
\expandafter\ifx\csname package@font\endcsname\relax\else
 \expandafter\expandafter
 \expandafter\usepackage
 \expandafter\expandafter
 \expandafter{\csname package@font\endcsname}% 
\fi
\begin{document}
\title{A dust particle based technique to measure potential profiles in a plasma}%
\author{Garima Arora}%
\email{garimagarora@gmail.com}
\author{P. Bandyopadhyay}
\author{ M.G. Hariprasad}
\author{A. Sen}
\affiliation{Institute For Plasma Research, HBNI, Bhat, Gandhinagar,Gujarat, India, 382428}%
\date{\today}
%**************************************************************
%#####################################################################################                  ABSTRACT
%************************************************************************************************
\begin{abstract}
A simple approach  to measure the potential profile in a plasma based on the visualization of charged tracer dust particles is reported. The method is used to experimentally determine the potential around a grounded wire that is 
mounted on the cathode of a DC glow discharge. Argon plasma is produced in a $\Pi$--shaped Dusty Plasma Experimental (DPEx) device. The tracer particles, consisting of a few micron sized mono-dispersive Melamine Formaldehyde (MF) grains, are made to flow over the grounded wire by suitable variations in the background gas flow. By a visual tracking of the individual particle trajectories, that yields their positions and velocities at various times, the potential values at these positions are directly estimated by using energy conservation arguments. The results agree very well with conventional probe based measurements. The technique is free of some of the inherent limitations of probe based diagnostics and offers a direct and minimally invasive means of visualizing potential profiles in a plasma. 
\end{abstract}
%%%%%%%%%%%%%%%%%%%%%%%%%%%%%%%%
\maketitle
%%%%%%%%%%%%%%%%%%%%%%%%%%%%%%%%%%%%%%%%%%%%%%%%%%%%%%%%%%%%%%%%%%%%%%%%%%%%%%%%%%%%%       INTRODUCTION
%%%%%%%%%%%%%%%%%%%%%%%%%%%%%%%%%%%%%%%%%%%%%%%%%%%%%%%%%%
\section{Introduction}\label{sec:intro}
Sheath formation is an ubiquitous phenomenon whenever a plasma is in contact with a material surface and these  fundamental potential structures have been extensively studied \cite{bohm,robertson,franklinsheath,sheathplasmafacingsurface} both theoretically and experimentally over the past few decades.  A sheath is basically a space charged region that develops between the physical boundary (e.g. the wall, electrodes \textit{etc}.) and the plasma, essentially because of the difference in the mobility of the electrons and ions that leads to a preferential negative charging of the wall. Despite the long history of research on sheaths there still remain many open questions relating to their structure and dynamics and the subject continues to receive active attention both at a fundamental level and in the context of its wide variety of applications.\\ 
Recently there has been a growing interest in dusty plasma flow experiments and the interaction of flows with potential structures in the plasma. A notable example is an experimental observation of precursor solitons \cite{surbhiPrecursor} in a dusty plasma medium caused by a supersonic flow of the dusty plasma over a stationary electrostatic potential hill. The shape and size of the potential hill was shown to determine the type of  nonlinear waves and structures that got excited. The experiment highlighted the need for precise measurements of the potential structures around such floating/biased electrodes in order to obtain a better understanding of the nonlinear excitation process and a more realistic theoretical interpretation of the same.\par  
A variety of electrostatic probes namely,  Langmuir probes \cite{merlinolangmuir,tonkslangmuir}, floating and laser heated emissive probes \cite{reviewarticleemisiveprobe,tonkslangmuir,goreesheath,experimentsheath,varalaserheatedemissive,laserheatedemissive} are commonly used to measure the space and plasma potential profiles even in the sheath region in a variety of plasmas.
The use of dust particles as microprobes in the plasma is another diagnostic technique that has been widely used in the past in various experiments to measure the electric field in the sheath region\cite{kersten2000,basner,samarian,hartmann,morfillfield,schubert2012,vladimirov,schubert2011,annaratone}. The equilibrium electric field \cite{kersten2000} and the potential structures in the sheath  are studied \cite{schubert2012} by looking at the dust particle's equilibrium position in the associated potentials and field structures. Dust particles are used as a probe to estimate the sheath profile in the radial direction of a RF discharge \cite{samarian,basner,annaratone}  by resonating the particle motion with an applied low frequency AC signal in the central pixel of a segmented adaptive electrode. Hartmann \textit{et al.} \cite{hartmann} rotated dust particles  and estimated the horizontal electric field by balancing the centrifugal force with the electrostatic force. The nature of these perturbation is studied by using suspended particles as tracers. In contrast to these single particle tracer studies, that are mainly aimed at diagnosing sheath potential structures in a plasma,  E. Thomas and collaborators have over the past few years used  the Particle Image Velocimetry (PIV) technique to study the internal dynamics of dust clouds as well as their interactions with high speed charged particle streams \cite{Thomas2001,Thomas2002a,Thomas2002b,Thomas2003,Thomas2004, Thomas2006,thomas}.  
However, for measurements of plasma potential structures such as sheath profiles it is more convenient to track a single dust particle which acts as a non-perturbative microprobe.  Most past studies using the microprobe methodology have relied on a series of static local measurements to map the profile of plasma potential structures. In our present work we employ the microprobe in a dynamical fashion by continuosly tracking the trajectory of a moving particle to provide a direct visualization of the potential profile. By launching the trace particle at different velocities it is possible to probe different regions of the potential structure. Our work is motivated by the need to know the size and shape of such potentials in flowing dusty plasma experiments that are aimed at exciting nonlinear wake structures \cite{surbhiPrecursor,surbhishock}. We demonstrate this method experimentally by measuring the potential profile around a grounded wire that is mounted on the cathode of a DC glow discharge plasma. Our experimental measurements carried out over a range of discharge parameters are found to be in good agreement with conventional measurements made with an emissive probe. Since the small tracer particles act like micro probes which essentially do not modify the potentials around the object, this method offers an accurate and minimally invasive means of measuring potentials in a plasma. \par
The paper is organized as follows: In the next section (Sec.~\ref{sec:setup}) a detailed description of the experimental set up along with the production procedure of the plasma and the introduction of dust tracer particles is given.  The experimental observations and results are discussed in Sec. ~\ref{sec:results}.  Sec.~\ref{sec:conclusion} provides a brief summary and discussion on our findings. \par
 %***********************************************************************************************************************
\section{Experimental Set-up}\label{sec:setup}
%%%%%%%%%%%%%%%%%%% FIGURE   %%%%%%%%%%%%%%%%%%%%%
\begin{figure}[ht]
\includegraphics[scale=0.47]{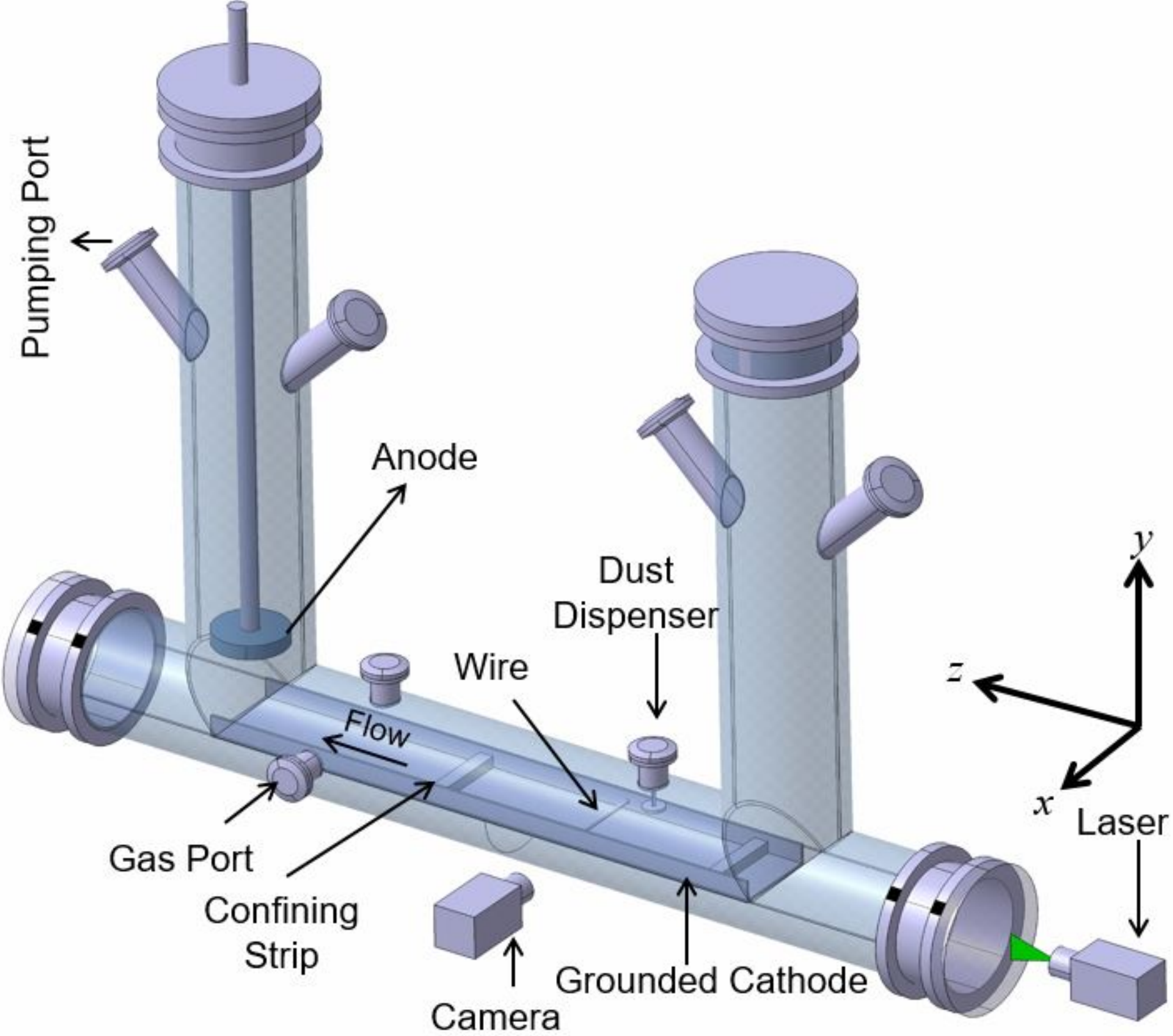}
\caption{\label{fig:fig1} A schematic diagram of dusty plasma experimental (DPEx) setup.  }
\end{figure}
%%%%%%%%%%%%%%%%%%%%%%%%%%%%%%%
 The experiments are performed in a $\Pi$-shaped Dusty Plasma Experimental (DPEx) device, whose schematic diagram is shown in Fig.~\ref{fig:fig1}. The experimental device has several radial and axial ports for the purpose of pumping, feeding gas, mounting electrodes, introducing the dust particles etc. The system geometry and its associated diagnostics have been explained in greater detail elsewhere \cite{surbhiRsi}. To produce the plasma, a stainless steel circular disc is used as an anode whereas a long grounded rectangular plate serves as a cathode. The edges of the cathode are bent to provide radial confinement of the charged micro particles and two additional strips are kept at a distance of 20 cm to provide axial confinement. A copper wire is mounted in between these two strips which acts as a stationary charged object and provides a potential barrier to the horizontal flow of the charged particles. This wire is always connected to ground throughout our experiments; however there is a provision to keep the wire in the floating potential or at an intermediate potential.\par
 %%%%%%%%%%%%%%%%%%% FIGURE   %%%%%%%%%%%%%%%%%%%%%
\begin{figure}[ht]
\includegraphics[scale=1.1]{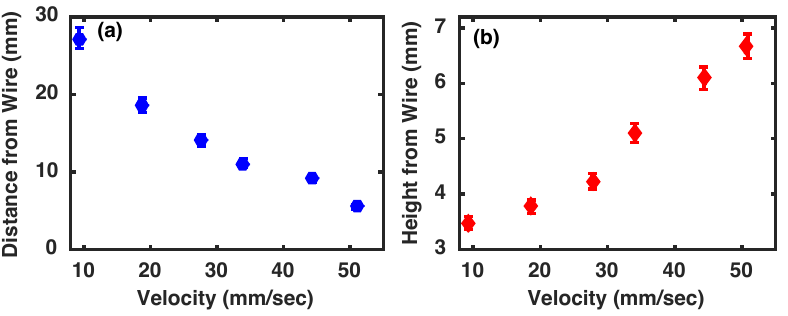}
\caption{\label{fig:fig2}  Variation of  (a) the closest approach and (b) the height achieved by a 4.45 $\mu$m particle above the charged wire versus the velocity of the micro particle. }
\end{figure}
%%%%%%%%%%%%%%%%%%%%%%%%%%%%%%%
%%%%%%%%%%%%%%%%%%% FIGURE   %%%%%%%%%%%%%%%%%%%%%
\begin{figure*}
\includegraphics[scale=0.35]{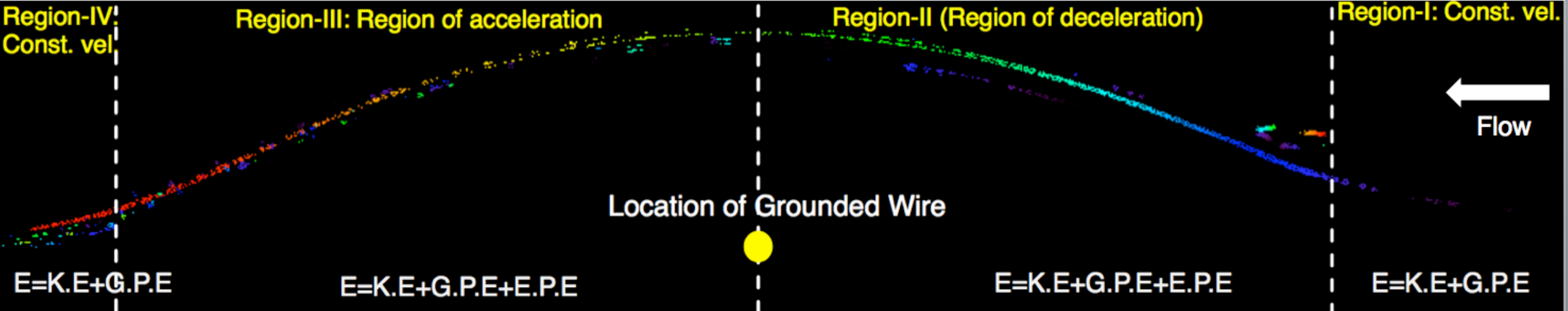}
\caption{\label{fig:fig3} Color plot showing the trajectory of the particle, having radius $a=4.45$ $\mu$m, while crossing the charged wire at a pressure of p=12 Pa and a discharge voltage $V_d=320$~V. The filled circle represents the location of the wire whereas the arrow represents the direction of flow.}
\end{figure*}
%%%%%%%%%%%%%%%%%%%%%%%%%%%%%%%
 To begin with, the experimental vessel is pumped down to a base pressure of $0.1$ Pa by a rotary pump. Argon gas is then flushed in the device several times with the help of a gas flow controller and every time the vacuum vessel is again pumped down to the base pressure to remove impurities from the vessel. Finally the working pressure is set to 10-20 Pa by adjusting the pumping rate and the gas flow rate. A plasma is then initiated by applying a voltage in the range of 280 to 360 V over which the plasma current varies from 1 to 10 mA.  With the creation of a plasma the wire mounted on the grounded cathode acquires a sheath around it. The measurement of the sheath potential structure is the main objective of our present experiment. \par
A few mono-dispersive MF particles of radius $a$ $=$ 4.45 $\mu$m and mass $m_d$ $=$ 6.1 $\times 10^{-13}$ kg are introduced into the plasma by shaking a dust dispenser. In the plasma environment these particles get negatively charged and levitate in the cathode sheath region. These micro particles levitate $\sim$ 4.5 mm above the wire. The exact height of the levitation depends on the discharge condition to attain the equilibrium. To illuminate these micro particles in the $y-z$ plane, a green laser is used. The Mie scattered light from the dust particles is captured by a CCD camera and recorded in a computer for analysis of the dust dynamics. \par
%%%%%%%%%%%%%%%%%%% FIGURE   %%%%%%%%%%%%%%%%%%%%%
\begin{figure}
\includegraphics[scale=0.42]{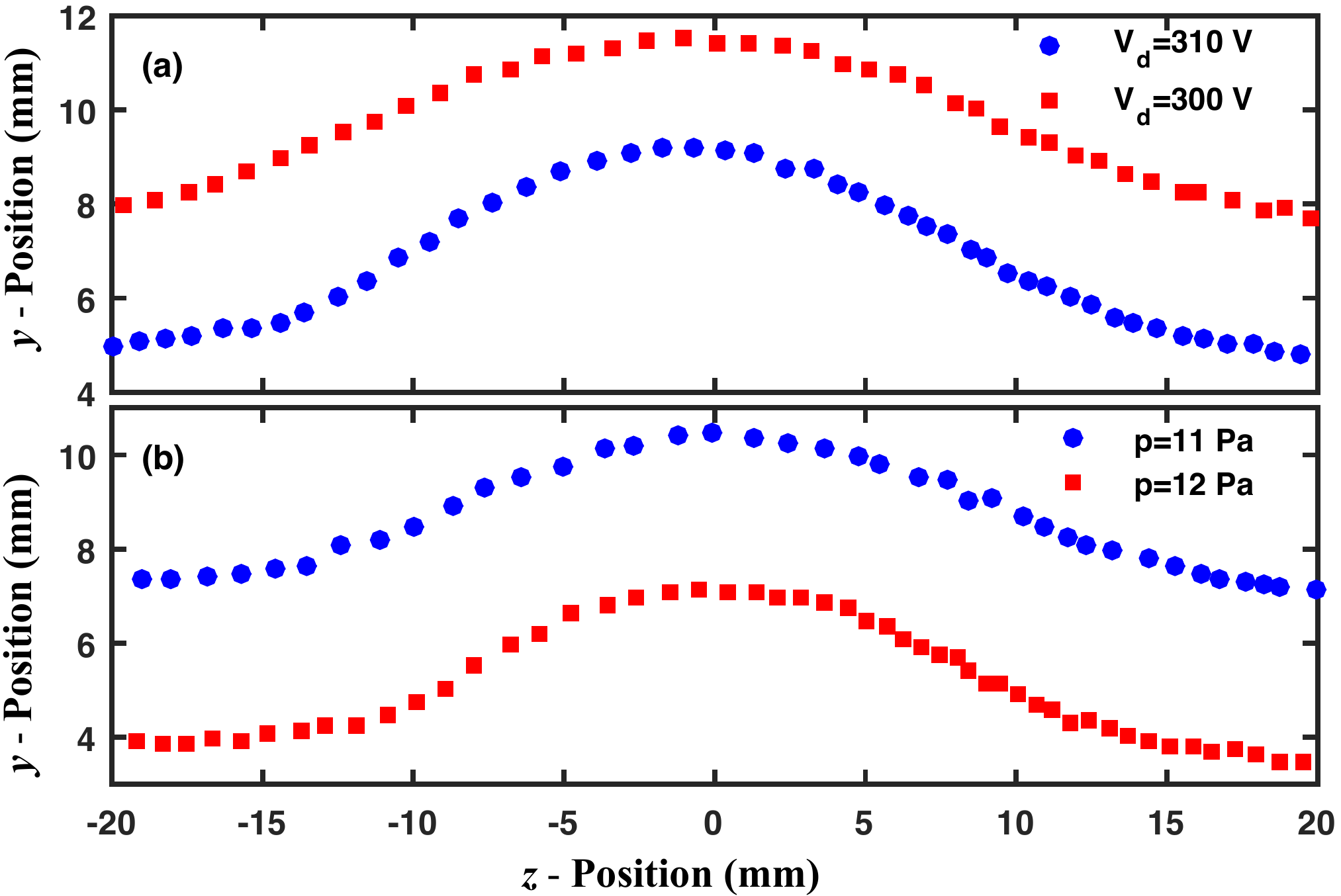}
\caption{\label{fig:fig4} A plot showing the trajectories of a particle of radius $a=4.45$ $\mu$m for different (a) discharge voltages and (b) background neutral pressures.}
\end{figure}
%%%%%%%%%%%%%%%%%%%%%%%%%%%%%%%
To initiate the flow of dust particles over the charged wire, the  single gas injection technique is used as discussed in detail in Ref. \cite{surbhiFlow}.  In this technique, an equilibrium condition of the dust particles is achieved at first by adjusting the pumping speed and the gas flow rate. The stationary dust particles are then made to flow by a sudden momentary decrease of the gas flow rate. During the flow, the particles are seen to move from right to left and they come back to their original position when the flow rate difference is set to original value. When the flow rate difference is increased beyond 5.5 sccm/min the particles move towards the pump with a higher velocity and cross the potential barrier created by the wire and finally fall down on the left edge of the glass tube where the cathode ends. The trajectory of the particle, as video recorded, is then used to find the strength and profile of the barrier produced by the wire. We discuss this in detail in the next section.
%%%%%%%%%%%%%%%%%%%%%%%%%%%%%%%% 
%%%%%%%%%%%%%%%%%%% FIGURE   %%%%%%%%%%%%%%%%%%%%%
\begin{figure*}[ht]
\includegraphics[scale=0.75]{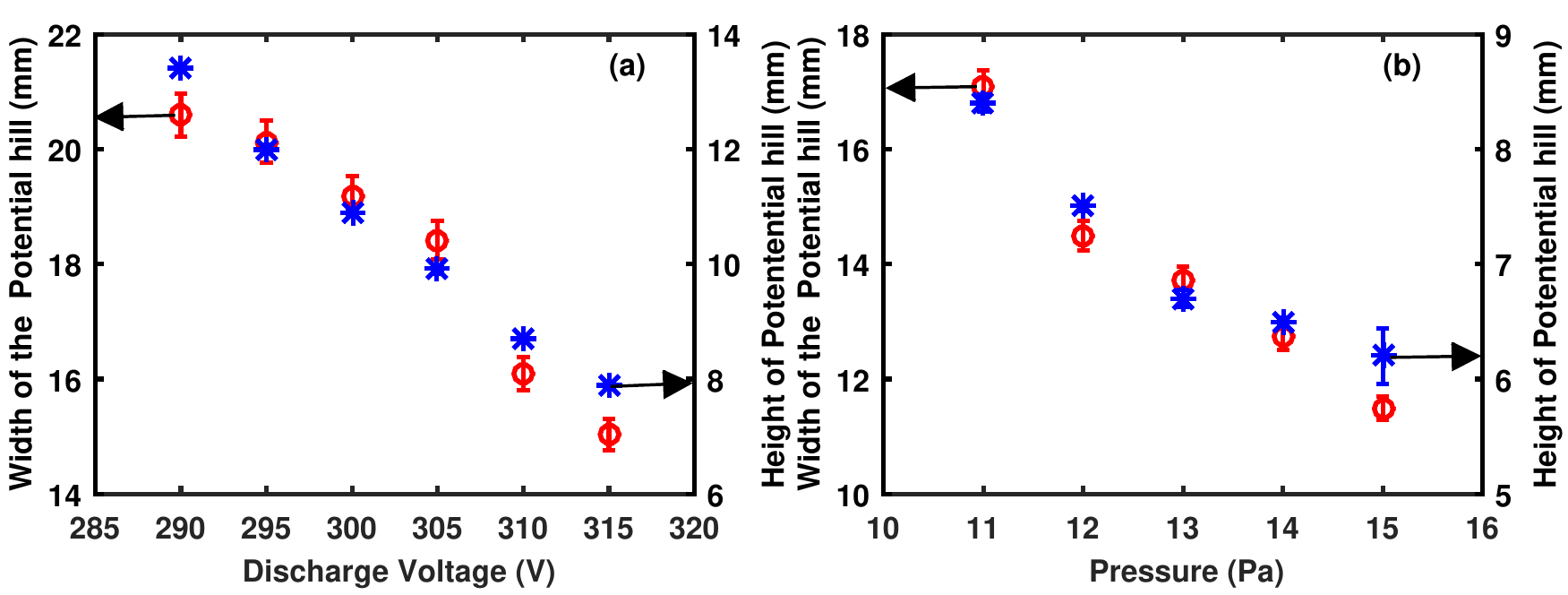}
\caption{\label{fig:fig5(a)} Variation of width (star) and height (open circle) of the potential hill created by the charged wire with (a) discharge voltages and (b) background neutral gas pressure.}
\end{figure*}
%%%%%%%%%%%%%%%%%%% FIGURE   %%%%%%%%%%%%%%%%%%%%%
\section{Results and Discussion}\label{sec:results}
As discussed above, the injected particles attain a stable equilibrium when the pumping rate gets exactly balanced by the gas flow rate. In this situation the particles are observed to be confined in the potential well between the right strip and the potential hill of the grounded wire. The particles display only small random displacements due to their thermal energy. Keeping the pumping speed constant when the gas flow rate is decreased, the particles are seen to move towards the pump (right to left) and approach the grounded wire. If the velocities of these particles are not sufficient to enable them to go over the potential hill, they stay inside the potential well but attain an equilibrium position closer to the wire. The variation of this closest distance of approach and the particle's height above the wire as a function of its velocity are shown in  Fig.~\ref{fig:fig2} for a gas pressure of $p=$ 12 Pa  and discharge voltage of $V_d=$ 320  V.  It is clear from Fig.~\ref{fig:fig2}(a) and Fig.~\ref{fig:fig2}(b) that the distance of closest approach in the axial direction decreases and the height of the particle increases with an increase of the particle velocity. In this particular condition, the closest distance from the potential hill  is seen to be $\sim$ 5 mm and the height achieved by the particle is $\sim$ 7 mm  for a velocity of  $\sim$ 5 cm/sec. With a further increase of the gas flow rate and hence an increase in the magnitude of the terminal velocity, the particles are found to overcome the confining potential and do not come back in the well. \par
Fig.~\ref{fig:fig3} depicts the trajectory of a particle while moving over the grounded wire. In this case, the terminal velocity crosses the threshold value ($\sim$ 6 cm/sec) so that it flows over the grounded wire and crosses the hill. The violet color (extreme right) of the trajectory indicates the particle coordinates at the initial time whereas the red color (extreme left) indicates the coordinates of the same particle at the final stage of its journey. The big solid circle shows the position of the charged wire. It is clear from the figure that the particle almost traces the sheath-plasma boundary (caused by the grounded potential wire) on its way. Very far from the potential wire, the particle moves with the terminal velocity (Region-I) towards the confining strip, climbs up (Region-II) the potential hill created by the wire and then moves down (Region-III) from the hill and finally (Region-IV) travels again  with almost the same terminal velocity that it started with. Using the coordinates of the individual particles we will now deduce the height and width of the potential hill and also delineate the axial and radial profiles of the potential structure. \par
Fig.~\ref{fig:fig4} displays the particle trajectories (similar to Fig.~\ref{fig:fig3}) for two different discharge voltages (Fig.~\ref{fig:fig4}(a)) and background neutral gas pressures (Fig.~\ref{fig:fig4}(b)) when all other discharge parameters are kept constant. It is seen that the height and the width of the potential hill increases with the decrease of the discharge voltages for a background pressure of $p=$ 12 Pa (see Fig.~\ref{fig:fig4}(a)). A similar trend is seen when the pressure is reduced for a particular discharge voltage $V_d=$ 310 V. It is also seen from Fig.~\ref{fig:fig4} that the $z$-coordinates of the particles are less spaced when the particles ride the hill near $\sim z=7$ mm whereas the coordinates are well spaced when the particles come down from the hill near $\sim z=-7$ mm. This essentially means that the particle decelerates while climbing up the hill whereas it gets accelerated when moving down the hill. Here the time interval of two consecutive frames is set to be $\sim$ 9.3 msec. \par
Fig.~\ref{fig:fig5(a)} shows the variation of the width and height of the potential hill created by the charged wire over a wide range of discharge parameters. It is seen from Fig.~\ref{fig:fig5(a)}(a), that both the width and the height of the hill decrease with the increase of the discharge voltage. A similar trend of the changes in the width and height is observed when the pressure is increased from 11 to 15 Pa as shown in Fig.~\ref{fig:fig5(a)}(b). In both the cases when the voltage or the pressure is increased, the plasma density increases which results in a decrease of the plasma Debye length. As the cathode sheath thickness is a function of the Debye length, hence the sheath thickness around the grounded potential wire decreases with the increase of discharge voltage and the background gas pressure. \par
To estimate the axial and radial potential profiles around the grounded object, we can make use of the energy conservation relation, 
\begin{eqnarray}
\left[\frac{1}{2}m_dv^2+m_dgh\right]_{I,IV}=\left[\frac{1}{2}m_dv^2 + m_dgh + Q\phi\right]_{II,III.}
\label{conserv}
\end{eqnarray}
\noindent
%%%%%%%%%%%%%%%%%%% FIGURE   %%%%%%%%%%%%%%%%%%%%%
\begin{figure}[ht]
\includegraphics[scale=0.75]{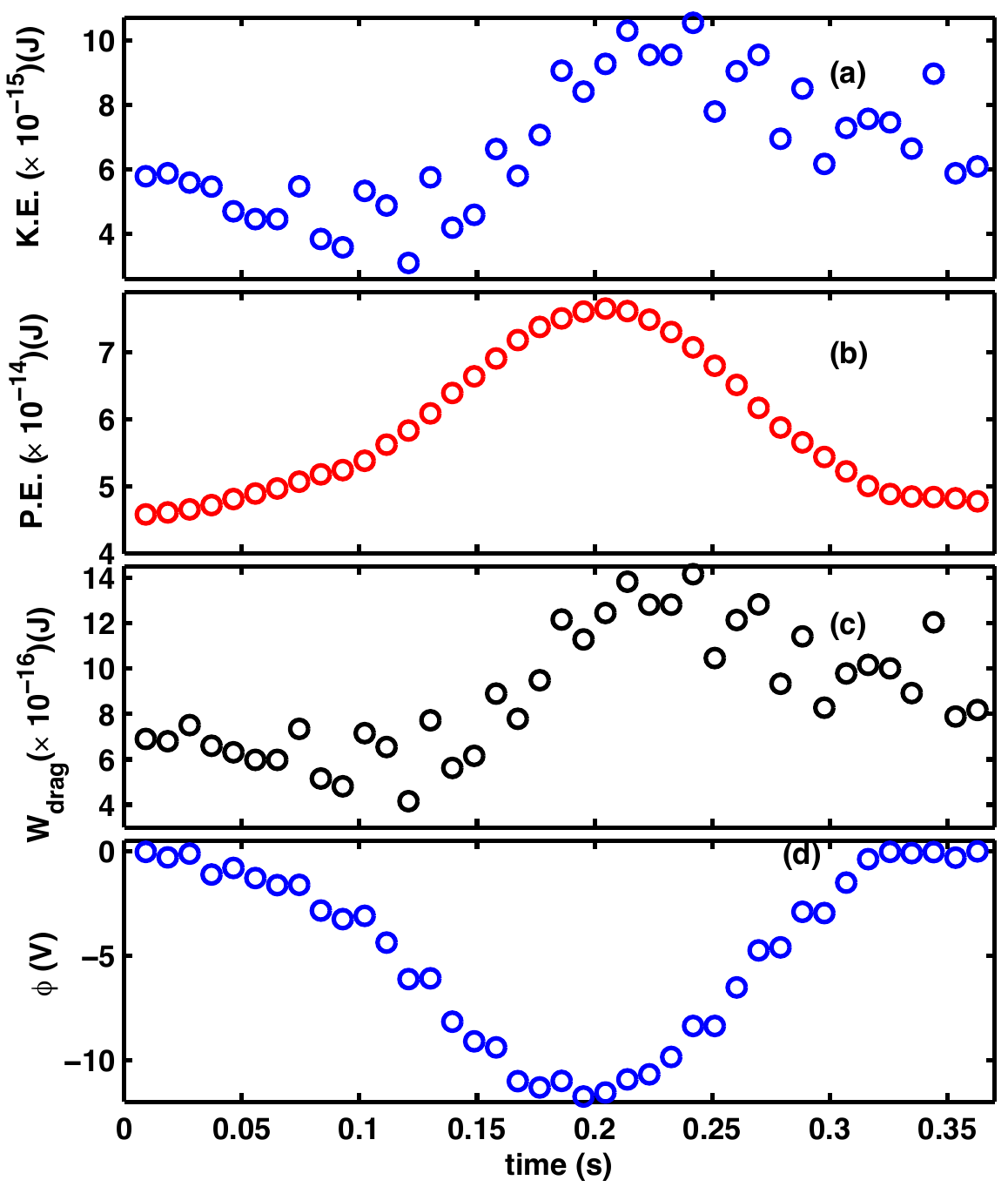}
\caption{\label{fig:fig6} Time evolution of (a) kinetic energy, (b) gravitational potential energy, (c) energy loss due to neutral drag force and (d) electric potential of a particle of radius $a=4.45$ $\mu$m while riding over a charged object for a given pressure  $p=12$ Pa and voltage $V_d=320$ V.}
\end{figure}
%%%%%%%%%%%%%%%%%%%%%%%%%%%%%%%
%%%%%%%%%%%%%%%%%%% FIGURE   %%%%%%%%%%%%%%%%%%%%%
\begin{figure*}[ht]
\includegraphics[scale=0.82]{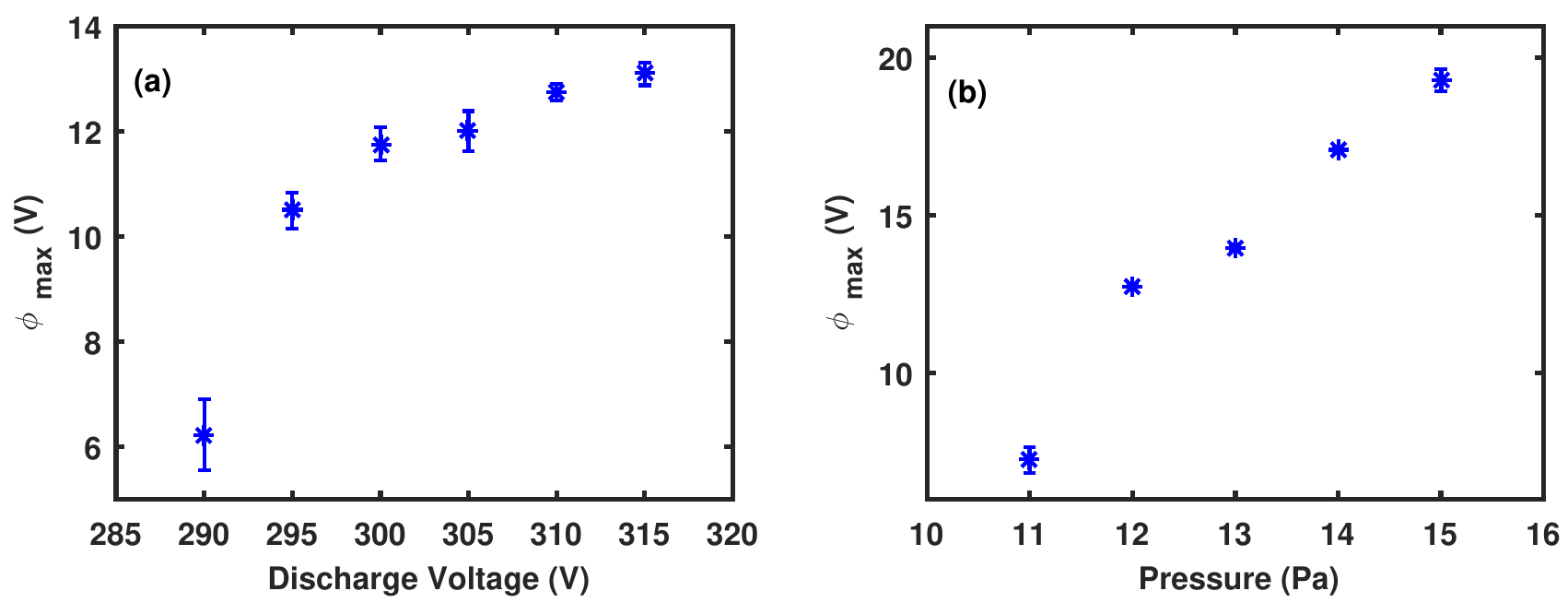}
\caption{\label{fig:fig7} Variation of maximum potential calculated from potential profiles for different discharge (a) voltages and (b) gas pressure.}
\end{figure*}
%%%%%%%%%%%%%%%%%%%%%%%%%%%%%%%
%%%%%%%%%%%%%%%%%%% FIGURE   %%%%%%%%%%%%%%%%%%%%%
\begin{figure}
\includegraphics[scale=0.78]{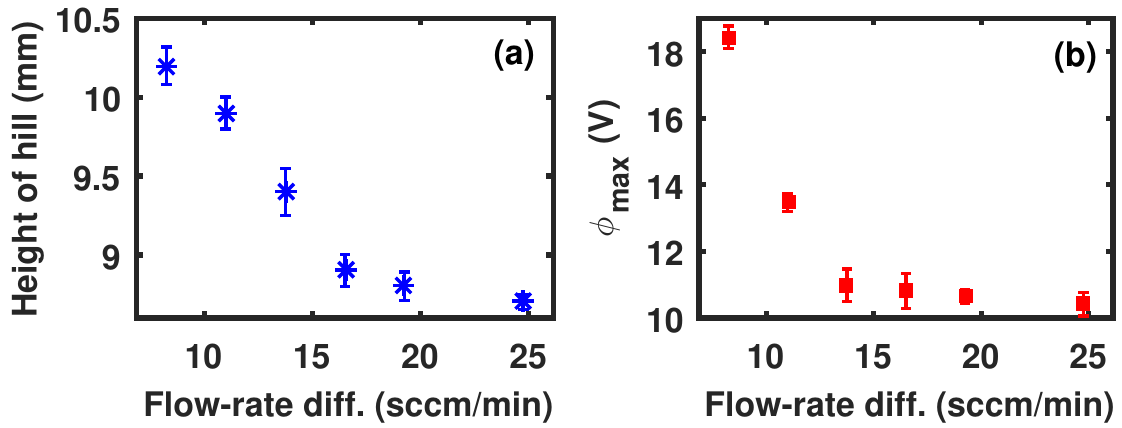}
\caption{\label{fig:fig8}  A plot showing the variation of (a) the height of the potential  with flow rate difference and (b) the maximum potential.}
\end{figure}
%%%%%%%%%%%%%%%%%%%%%%%%%%%%%%%
where, $m_d$, $v$, and $h$ denote the mass, velocity and height of the particles (above the cathode tray), respectively. $\phi$ is the electrostatic potential created by the grounded wire, $g$ is the acceleration due to gravity and $Q $($=-Z_de$ = $4\pi\epsilon_0a\phi_s$), where $Z_d$ is dust charge number, $e$ is the electronic charge and $\phi_s$ is the dust surface potential for an Argon plasma (which is negative in laboratory dusty plasmas), $\epsilon_0$ is the permittivity of free space, $a$ is the radius of the dust particle. The subscripts I,II,III,IV refer to the various regions of the trajectories as indicated in Fig.~(\ref{fig:fig3}). As shown in Fig.~\ref{fig:fig3}, when the particles stay far from the wire (e.g., Region-I and Region-IV) they  move only with the constant kinetic energy (K.E)) and the gravitational potential energy (G.P.E). But when they enter in the sheath region, the dynamics of the particles get changed and they start feeling the presence of the charged wire. Hence in Region-II and Region-III, the components of total energy are kinetic energy, gravitational potential energy and electrostatic potential energy (E.P.E). As a result the potential profile $\phi(z,y)$ created by the wire can be expressed as:
 \begin{eqnarray}
 \phi(z,y)=\frac{m_d}{Q} \left[  \left[\frac{1}{2}v_d^2+gh\right]_{Z\rightarrow\infty} -\frac{1}{2}v^2 -gh \right] 
 \label{conserv2} 
\end{eqnarray}
%%%%%%%%%%%%%%%%%%% FIGURE   %%%%%%%%%%%%%%%%%%%%%
\begin{figure*}
\includegraphics[scale=0.7]{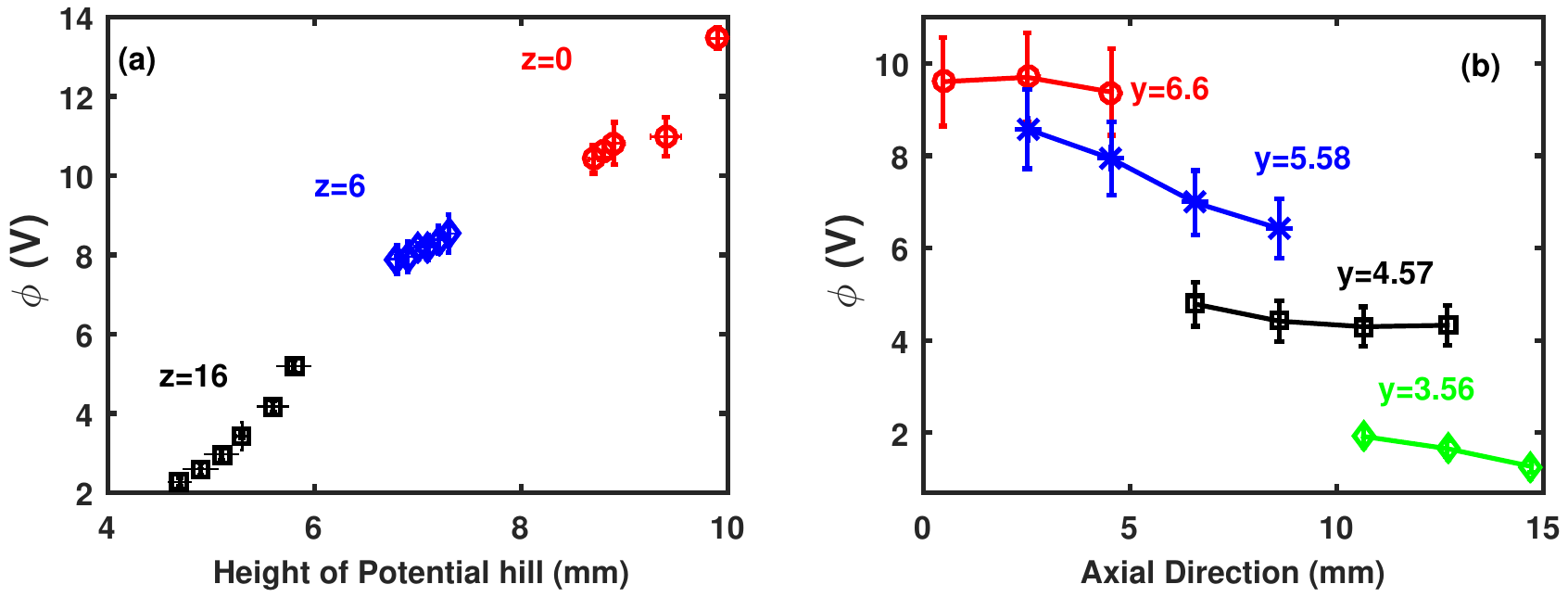}
\caption{\label{fig:fig9}(a)  Radial potential profiles at z=0, z= 6 mm and z=16 mm.  z=0 corresponds to the location of the wire. and (b) Axial potential profiles at y=6.6, y=5.58~mm, y=4.57~mm, y=3.56~mm}
\end{figure*}
%%%%%%%%%%%%%%%%%%%%%%%%%%%%%%%
where $v=\sqrt {v_{z}^2+v_y^2}$ and $h$ is the height at a particular axial location $z$ and $v_d$ is the terminal velocity in Region-I.
By tracking individual particles, the potential profile $\phi (z,y)$ along the trajectory can be estimated with the help of Eq. (\ref{conserv2}). \par 
For determining the potential profile, we first need to estimate the charge acquired by the dust particle in equilibrium. To estimate the dust charge, a Collision Enhanced plasma Collection (CEC) model \cite{khrapak1,khrapak2} has been used to first calculate the surface potential, $\phi_s\sim -\tilde{z}K_BT_e/e$ of the dust particle by solving the electron and ion flux balance equations numerically and subsequently an estimate of the charge is made using $Q=4\pi\epsilon_0a\phi_s$  assuming that the dust particles are spherical in shape. For our experimental conditions  $p=11-15$ Pa, $T_e=4-2.5$ eV and $n_i=0.9-2 \times 10^{15}$ $/m^3$ \cite{surabhiRsi},  $\tilde{z}$ and $Q$ come out to be $\sim 1.34-1.56 $ and $\sim 2.6-1.9 \times10^{-15}$ C, respectively. It is to be noted that for the basic plasma parameters such as $T_e$ and $n_i$, we have used values previously obtained from actual measurements using a single Langmuir probe for discharges of similar conditions. These have been discussed and reported in \cite{surabhiRsi}.
Here, we have assumed that the charge of the dust particle is almost constant and does not change during its journey. This assumption is based on findings from earlier experiments carried out in the same device and for similar conditions \cite{surbhiRsi}. These experiments have shown that the plasma parameters remain nearly constant along the axial direction (along $z$). In particular, the constancy of $T_e$ ensures that the charge remains almost constant \cite{merlinocharge,goreecharging}. When these micro-probes are made to flow over a grounded wire, their kinetic energy and potential energy profiles are measured and it is seen that the particles decelerate while riding up the potential hill and they accelerate while  coming down from the hill (see Fig.~\ref{fig:fig6}(a)). Fig.~\ref{fig:fig6}(b) indicates that the particle starts its journey at a height of $\sim$~4.5 mm from the wire and then attains a maximum height and then ends its journey almost at the same height that it started from. It is to be noted that the energy loss due to dust neutral friction is not taken into account in Eq. (\ref{conserv}). This is a reasonable assumption as it is seen experimentally that the particles display the same kinetic energy and height after climbing down from the potential hill indicating that there is negligible energy loss due to collisional friction with the neutrals. This is a consequence of the experiments being carried out in a low pressure regime.  However, in a higher pressure range, where collisional effects may be important, one should consider the instantaneous energy loss due to dust-neutral collisions. This drag force is calculated as
$W_{drag} = \int_{s(t_1)}^{s(t_2)}{\vec{F}_n.d\vec{s}}$ where $F_n = N m_n\pi a^2v_{tn}v_g$ \cite{Thomas2001} and $s(t_1)$, $s(t_2)$ are the positions of the particle at times $t_1$ and $t_2$ along its trajectory.
$N$, $m_n$, $a$, $v_{tn}$, $v_g$ are the number density of neutrals, mass of neutral, radius of dust particle, thermal velocity of neutrals and dust grain velocity, respectively. For our experimental conditions this loss varies from $\sim 4.2-14.2 \times10^{-16}$ J as shown in the Fig.~\ref{fig:fig6}(c) which is nearly one order of magnitude smaller than the kinetic energy and two orders of magnitude than the gravitational potential energy and hence its neglect is justified. With the help of the kinetic and gravitational potential energy profiles, the electrostatic potential profile is estimated using Eq. (\ref{conserv2}) and is shown in Fig.~\ref{fig:fig6}(d). The electrostatic potential profile follows an almost a symmetric profile around the wire and a similar kind of potential profile was obtained when the floating/plasma potentials were measured using Langmuir and emissive probes \cite{surbhiRsi}. \par
Fig.~\ref{fig:fig7} shows the variation of the maximum potential strength (near the location of the wire) created by the wire with a variation of the discharge parameters.  The strength of the potential is seen to increase with an increase in the discharge voltage for a constant neutral gas pressure $p=12$ Pa (Fig.~\ref{fig:fig7}(a)) or alternatively with an increase in the neutral pressure keeping the discharge voltage fixed at $V_d=$ 310 V (Fig.~\ref{fig:fig7}(b)). A similar trend was observed earlier in experiments reported by Jaiswal \textit{et al}.\cite{surbhiRsi}
To fully explore the radial and axial potential profiles around the grounded wire, we have varied the difference of gas flow rate (and hence the terminal velocity) over a wide range such that in all the cases the particles can overcome the potential barrier. In each case, the particles gain different terminal velocities and as a result in order to satisfy the conservation of total energy, they cross the potential hill at different heights. It is observed that the height of the potential hill decreases when the difference of gas flow rate increases as shown in Fig.~\ref{fig:fig8}(a). This figure is plotted for a constant gas pressure of $p=12$ Pa and a constant discharge voltage of $V_d$=320 V. The strength of the potential, estimated from the conservation of energy, at the peak of the hill is shown in figure Fig.~\ref{fig:fig8}(b). It clearly indicates that for different values of flow rate difference the strength of potential decreases with height. \par
 To construct the radial  potential profile, we plot the variation of potential strength with height as shown in Fig.~\ref{fig:fig9}. Fig.~\ref{fig:fig9}(a) shows the radial potential variation for three different axial locations $z=0$ mm, $6$ mm and $16$ mm respectively, where $z=0$ mm corresponds to the location of the wire. It is to be noted that Fig.~\ref{fig:fig8}(a) and Fig.~\ref{fig:fig8}(b) have been used to plot Fig.~\ref{fig:fig9}(a). A similar exercise is done to construct the potential for other  two axial locations. Fig.~\ref{fig:fig9}(a) shows that the  magnitude of the potential increases in all the cases when one goes away from the sheath created by the grounded wire towards the bulk plasma in radial direction. However, the potential rise is maximum at the location of the wire as compared to the other axial locations.  Fig.~\ref{fig:fig9}(b) shows the variation of potential strength in axial direction for four different radial locations $y=6.6$ mm, $y=5.58$ mm, $y=4.57$ mm and $y=3.56$ mm. It is clear from this figure that the variation of potential is not significant as we move away from the wire axially for a particular radial location.\par
Finally we would like to remark that our experimental findings have great relevance for past and future dust flow experiments aimed at excitation of linear and nonlinear wave structures.  In the past experiment of Jaiswal \textit{et al.} \cite{surbhiPrecursor}, precursor solitons were excited by a supersonic mass flow of the dust particles over a stationary electrostatic potential hill. A theoretical explanation of these excitations based on the forced Korteweg-deVries (fKdV) model equation was provided by assuming the shape of the potential hill to be Gaussian and using arbitrary values of the amplitude and width of this potential to solve the model equation. By doing so they were only able to provide a qualitative comparison between the experimental and theoretical results. Our present results not only confirm the shape of the potential to be close to a Gaussian but provide experimental measures of the amplitude and width of the potential for carrying out a quantitative comparison with modeling results. Furthermore, our measurements using the particle tracing technique also show that the amplitude and width of the potential can change with the discharge voltage and the background pressure as displayed in Fig.~5 and Fig.~6 respectively. These results offer valuable clues for further means of manipulating the size of the potential and can prove useful for planning future experiments on flow induced nonlinear excitations  over a wider range of parameter space.
\section{conclusion}
\label{sec:conclusion}
In conclusion, we have experimentally demonstrated a simple technique to measure the potential around a charged object using the flow of a few tracer dust particles. The experiments have been carried out in the DPEx device in which an Argon discharge was initiated between a disc shaped anode and a grounded long cathode. A few micron sized dust particles were then introduced into the plasma, which were found to float near the cathode sheath boundary. A steady state equilibrium of the dust particles was achieved by a fine tuning of the  pumping speed and gas flow rate. A flow of the dust particles was then initiated by a sudden decrease of the mass flow rate of neutral gas. As a result the particles traveled  over the grounded wire which was placed on the path of the particles. From a knowledge of the instantaneous position of the particles and their velocities, their potential and kinetic energies were estimated over a wide range of discharge parameters. The axial and radial electrostatic potential around the charged object was estimated by employing the conservation of total energy. In this technique, the particles act as dynamic micro-probes and measure the potential very precisely without significantly perturbing  the potential around the charge object.\par 
Our experimental results show that the potential around a small charged object has a symmetric profile in the axial direction whereas it has a parabolic shape in the radial direction and the dimension of the potential shrinks in all directions with an increase of the discharge voltage and the background neutral gas pressure.
In our experiments, we have used a grounded wire as a sample problem to estimate the potential profile around it. It does not mean that the method is restricted to such a profile.  For any shape of the potential, the instantaneous height and the velocity of the particle will change and accordingly provide us with values of the potential at different points.\par
The presence of plasma homogeneity is a major simplifying factor in our experiments. However its absence does not invalidate our approach. Inhomogeneity would only change the magnitude of the charge of dust particles (due to the change of plasma parameters) at different locations. The potential, in that case, can still be measured by following the same technique with the known value of charge.
Thus this technique can not only be implemented in such devices \cite{pk41,merlino,nakamura,surbhiRsi,bailung}  whose electrode (plasma) configurations are akin to ours but also in devices that have more complex electrode configurations. \par
We believe that this simple technique has potential advantages over conventional active diagnostic tools like probes in flowing dusty plasma experiments as a non-invasive diagnostic tool. Like other diagnostics, this technique also has some disadvantages. To measure the potential, one has to initiate a flow of the dust particles with respect to the background plasma. In addition, this technique allows us to measure the potential only at those points through which the particle travels. Nevertheless it does provide a simple and visual approach that can find useful applications in a variety of laboratory and plasma based industrial devices where sheath dynamics can play an important role. 
\section{Acknowledgements}
\label{sec:ACKNOWLEDGEMENTS}
A.S. thanks the Indian National Science Academy (INSA) for their support under the INSA Senior Scientist Fellowship
scheme.
%\bibliography{bib_file_final}
%merlin.mbs aipnum4-1.bst 2010-07-25 4.21a (PWD, AO, DPC) hacked
%Control: key (0)
%Control: author (8) initials jnrlst
%Control: editor formatted (1) identically to author
%Control: production of article title (0) allowed
%Control: page (1) range
%Control: year (1) truncated
%Control: production of eprint (0) enabled
%

\end{document}